\documentclass[11pt]{article}
\usepackage[a4paper,margin=24mm]{geometry}
\usepackage{amsmath,amssymb,amsthm,mathtools,microtype}
\usepackage[colorlinks=true,allcolors=blue]{hyperref}
\hypersetup{pdftitle={Time-invariant negativity and persistent bound entanglement in a cyclic two-qutrit family},pdfauthor={Mazhar Ali; Haobo Ma; Wenlin Zhang}}
\newtheorem{theorem}{Theorem}

\newtheorem{proposition}{Proposition}
\newcommand{\ket}[1]{|#1\rangle}
\newcommand{\bra}[1]{\langle#1|}
\newcommand{\Tr}{\operatorname{Tr}}
\newcommand{\id}{\mathrm{id}}
\newcommand{\R}{\mathcal R}
\newcommand{\spec}{\operatorname{spec}}
\title{Time-invariant negativity and persistent bound entanglement in a cyclic two-qutrit family}
\author{Wenlin Zhang$^{1,2}$, \qquad Mazhar Ali$^3$ \qquad Haobo Ma$^2$} 
\date{\small
\begin{minipage}{0.96\textwidth}\centering
$^1$National University of Singapore, 117417 Singapore\\[5pt]
$^2$ChronoAI Pte Ltd, Singapore\\
$^3$Department of Electrical Engineering, Faculty of Engineering,\\
Islamic University of Madinah, 107 Madinah, Saudi Arabia\\[3pt]
\href{mailto:e1327962@u.nus.edu}{e1327962@u.nus.edu}\\[3pt]
\href{mailto:mazhar.ali@iu.edu.sa}{mazhar.ali@iu.edu.sa}\quad
\href{mailto:auric@aelf.io}{auric@aelf.io}\\
W. Zhang: \href{https://orcid.org/0009-0005-2414-7968}{ORCID}\qquad
M. Ali: \href{https://orcid.org/0000-0002-3708-1970}{ORCID}\qquad
H. Ma: \href{https://orcid.org/0009-0008-4944-977X}{ORCID}
\end{minipage}\\[10pt]31 October 2026}
\begin{document}
\maketitle
\begin{abstract}
We study the entanglement dynamics of two qutrits under collective dephasing for a family of states obtained by mixing a maximally entangled state that lives in the decoherence-free subspace with three nonstationary two-term states sharing a common amplitude ratio $r$. We obtain exact PPT and separability regions along the whole trajectory, a specialization of known results on cyclic sign invariant states, and we supply explicit certificates: a finite product decomposition, three fixed positive-map witnesses, and a closed realignment formula that detects every PPT-entangled state, so that no state is left undecided. The negativity is computed in closed form. It shows that the entanglement is time-invariant, positive and constant from the initial time although the state changes, exactly when the weight of the maximally entangled state exceeds the threshold $U_r=r/(1+r+r^2)$. To our knowledge, this is the first demonstration of time-invariant entanglement for qutrit--qutrit systems under collective dephasing. For smaller weights the entanglement decays and vanishes at a finite time: the state either becomes separable (entanglement sudden death) or, for $r\neq1$ in an intermediate range, loses distillability suddenly and remains bound entangled at all later times. Every state with negative partial transpose is one-copy distillable by a fixed local projection. The same trajectories arise from the global dephasing map and from a correlated local random-unitary semigroup, although the two channels differ on product inputs. Exact rational benchmarks and embeddings of the Horodecki and Ha--Kye--Park families are provided.
\end{abstract}

\section{Introduction}
Quantum entanglement has been widely recognized as non-classical correlation with growing number of applications in the future technologies \cite{Erhard-NRP-2020,Friis-2019}. Due to its tremendous importance for technologies as well its fundamental role in our understanding of nature, it is essential to study detection and characterization of quantum entanglement \cite{Horodecki-RMP81,Guehne-PR474}. Many authors have studied this challenging task for both bipartite as well multipartite quantum states \cite{Horodecki-PLA223,Rudolph,ChenWu,Yu-PRL127,Imai-PRL126,Ali-QIP22}. For bipartite systems, Peres-Horodecki theorem is a key result which simply says that the partial transpose of a separable quantum state is positive (PPT) \cite{Peres}. For qubit-qubit and qubit-qutrit systems, PPT test is necessary and sufficient criterion to determine the separability/entanglement of a quantum state. However, for dimentions higher than $6$, there is clear evidence that entangled states can be PPT \cite{HorodeckiBound}. PPT-entangled states are called bound-entangled states as well, because their entanglement can not be distilled with local operations and classical communication (LOCC) \cite{Bennet-PRL82}. Several interesting families of bound-entangled states have been identified for different dimensions of Hilbert space along with their detecttion methods \cite{Horodecki,Bennet-PRL82,QiHou,HaKyePark}. It is of utmost importance to identify new families of bound-entangled states. In particular, if there are more families of bound-entangled states for a given dimension of Hilbert space, then it is interesting to study their relationship with each other.  

As theoretical studies and predictions about entanglement are rapidly brought to experimental verification, we also need to study the detrimental effects of environment on entanglement. Many studies have explored decoherence effects on entanglement for both bipartite and multipartite quantum systems \cite{Yu-work,Yu-work-2,Yu-work-3,lifetime,lifetime-2,Aolita-PRL100-2008,bipartitedec,bipartitedec-2,bipartitedec-3,Band-PRA72-2005,Band-PRA72-2005-2,Band-PRA72-2005-3,lowerbounds,Lastra-PRA75-2007,Guehne-PRA78-2008,Lopez-PRL101-2008,Ali-work,Ali-work-2,Ali-work-3,Weinstein-PRA85-2012,Weinstein-PRA85-2012-2,Ali-JPB-2014,Ali-JPB-2014-2,Ali-JPB-2014-3,Ali-JPB-2014-4,Ali-2015,Ali-2015-2,Ali-2016,Ali-2017,Ali-2017-2}. One specific noise dominant in ion-trap experiments is due to intensity fluctuations of electromagnetic fields leading to collective dephasing. The effects of collective dephasing on dynamics of entanglement have been studied for both bipartite and multipartite quantum 
systems \cite{Yu-work,Yu-work-2,Yu-CD-2002-3,AJ-JMO-2007,Li-EPJD-2007,Karpat-PLA375-2011,Liu-arXiv,Carnio-PRL-2015,Carnio-NJP-2016,Song2009,Ali2010}. Two interesting features of entanglement dynamics were observed: one so called {\it freezing dnamics} \cite{Ali-2017,Ali-2017-2} and {\it time-invariant dynamics} \cite{Karpat-PLA375-2011,Liu-arXiv,Carnio-PRL-2015,Carnio-NJP-2016}. In freezing dynamics, entanglement may first decay to some fixed value and then maintain this value despite quantum states change with time \cite{Carnio-PRL-2015}, whereas time-invariant dynamics refer to situations, where quantum states undergo dynamical process but their entanglement is always stationary from start to end. This feature was first observed for 
qubit-qutrit systems \cite{Karpat-PLA375-2011} and later for qubit-qubit systems \cite{Liu-arXiv}. In our earlier study, we have investigated both time-invariant dynamics and freezing dynamics of entanglement for qutrit-qutrit ($3 \otimes 3$) systems \cite{Ali-MPLA-2019}. The initial states considered in that study exhibited freezing dynamics, but no time-invariant entanglement was found. The present family gives explicit nonstationary two-qutrit trajectories with strictly positive, time-independent negativity.  

Section 2 introduces the dephasing map and the common-ratio family, proves its complete PPT and separability classification, and derives the negativity and transition times. Section 3 identifies known static families and gives exact dynamical examples. Section 4 summarizes the consequences and remaining extensions.

\section{Collective dephasing for qutrit-qutrit systems and the common-ratio family} 
\label{Sec:Model}

Let us consider two qutrits (two three-level atoms for example) $A$, $B$ coupled to collective dephasing noise. We take qutrits to be sufficiently far apart from each other and can be treated as independent. We choose the computational basis 
$ \{ \, |00\rangle$, $|01\rangle$, $|02\rangle \, |10\rangle$, $|11\rangle$, $|12\rangle$, $|20\rangle$, $|21\rangle$, and $|22\rangle$ $\}$. The time-dependent density matrix for two-qutrits system can be written \cite{AJ-JMO-2007} as  
\begin{eqnarray}
\rho(t) = \, \sum_{k=1}^3 \, (D_k^{\dagger} \, \rho(0) \, D_k)\, , \label{Eq:GS}
\end{eqnarray}
where   
$D_1 = \mathrm{diag}$ $(\gamma(t), 1,1,1,\gamma(t) \,,1,1,1,\gamma(t))$, 
$D_2 = \mathrm{diag} (\omega_1(t), 0,0,0,\omega_2(t) \,,0,0,0,\omega_2(t))$, and
$D_3 = \mathrm{diag} (0,0,0,0,\omega_3(t),0,0,0,\omega_3(t))$. $D_j$ are Kraus operators satisfying $\sum_j^3 \, D_j^\dagger D_j = \mathbb{I}$. The parameters are defined as, $\gamma = \mathrm{e}^{-\Gamma t/2}$, $\omega_1(t) = \sqrt{1-\gamma^2(t)}$, 
$\omega_2(t) = - \gamma^2(t) \, \sqrt{1-\gamma^2(t)}$, and $\omega_3(t) = (1-\gamma^2(t)) \, \sqrt{1+\gamma^2(t)}$.

The matrix form of Eq.~(\ref{Eq:GS}) for an arbitrary initial state is given as
\begin{equation}
\resizebox{0.96\textwidth}{!}{$\displaystyle
\rho(t) = \left(
\begin{array}{lllllllll}
\rho_{11} & \gamma \rho_{12} & \gamma \rho_{13} & \gamma \rho_{14} & \gamma^4 \rho_{15} & \gamma \rho_{16} & \gamma \rho_{17} & \gamma \rho_{18} & \gamma ^4 \rho_{19} \\
\gamma \rho_{21} & \rho_{22} & \rho_{23} & \rho_{24} & \gamma \rho_{25} & \rho_{26} & \rho_{27} & \rho_{28} & \gamma \rho_{29} \\
\gamma \rho_{31} & \rho_{32} & \rho_{33} & \rho_{34} & \gamma \rho_{35} & \rho_{36} & \rho_{37} & \rho_{38} & \gamma \rho_{39} \\
\gamma \rho_{41} & \rho_{42} & \rho_{43} & \rho_{44} & \gamma \rho_{45} & \rho_{46} & \rho_{47} & \rho_{48} & \gamma \rho_{49} \\
\gamma^4 \rho_{51} & \gamma \rho_{52} & \gamma \rho_{53} & \gamma \rho_{54} & \rho_{55} & \gamma \rho_{56} & \gamma \rho_{57} & \gamma \rho_{58} & \rho_{59} \\
\gamma \rho_{61} & \rho_{62} & \rho_{63} & \rho_{64} & \gamma \rho_{65} & \rho_{66} & \rho_{67} & \rho_{68} & \gamma \rho_{69} \\
\gamma \rho_{71} & \rho_{72} & \rho_{73} & \rho_{74} & \gamma \rho_{75} & \rho_{76} & \rho_{77} & \rho_{78} & \gamma \rho_{79} \\
\gamma \rho_{81} & \rho_{82} & \rho_{83} & \rho_{84} & \gamma \rho_{85} & \rho_{86} & \rho_{87} & \rho_{88} & \gamma \rho_{89} \\
\gamma ^4 \rho_{91} & \gamma \rho_{92} & \gamma \rho_{93} & \gamma \rho_{94} & \rho_{95} & \gamma \rho_{96} & \gamma \rho_{97} & \gamma \rho_{98} & \rho_{99}
\end{array}
\right) \,.$}\label{Eq:MF}
\end{equation}
Decoherence free subspaces (DFS) \cite{Yu-work,Yu-work-2,Yu-CD-2002-3} are common characteristic of collective dephasing. We also note that matrix elements $\rho_{15}$ and $\rho_{19}$ decay much faster as compared with all others. Finite times have $0<\gamma \le1$ and $\gamma=0$ denotes the asymptotic limit. Equivalently, Eq.~\eqref{Eq:MF} is the entrywise map $\rho_{uv}\mapsto\gamma^{(h_u-h_v)^2}\rho_{uv}$ with the charge array $h=(1,0,0,0,-1,0,0,0,-1)$ in the above basis order.

We construct a family of quantum states by mixing maximally entangled state living in DFS and another entangled state as defined below: Ali~\cite{Ali-MPLA-2019} studied freezing dynamics of entanglement in this two-qutrit dephasing with no trace of time-invariant dynamics for his choice of initial quantum states. Motivated by this setting, we consider a stationary maximally entangled ket mixed with three nonstationary two-term kets sharing an amplitude ratio $r>0$. The symmetric slice $r=1$ has time-invariant positive negativity but no PPT entanglement. An imbalance $r\ne1$ permits a persistent PPT-entangled tail after the negativity vanishes.
\begin{equation}
\begin{gathered}
 \ket{\eta}=(\ket{02}+\ket{10}+\ket{21})/\sqrt3,\\
\ket{ \chi_0(r)}=\frac{r\ket{00}+\ket{12}}{\sqrt{1+r^2}},\qquad
\ket{ \chi_1(r)}=\frac{r\ket{11}+\ket{20}}{\sqrt{1+r^2}},\qquad
\ket{ \chi_2(r)}=\frac{r\ket{22}+\ket{01}}{\sqrt{1+r^2}},
\end{gathered}
\label{eq:kets}
\end{equation}
and define, for $0\le p\le1$ and $r>0$, 
\begin{equation}
\rho_{p,r} = p \, \ket\eta\bra\eta +  \frac{1-p}{3} \, \sum_{j=0}^2\ket{\chi_j(r)}\bra{\chi_j(r)} \,.
\label{eq:mf}
\end{equation}
As we analyze below, these states are quite interesting. At the initial time $\gamma=1$ and $r = 1$, the states are NPT for all possible range of parameter $p$ except at $p = 1/3$, where we demonstrate that it is separable. It is interesting to see that for $r \ne 1$, some of the PPT-region has bound entangled states. The time evolution of these states under collective dephasing is simply written as
\begingroup
\begin{equation}
\begin{aligned}
 \rho_{p,r}(\gamma)
 &=\mathcal E_\gamma\left(p\ket\eta\bra\eta+
 \frac{1-p}{3}\sum_{j=0}^2\ket{\chi_j(r)}\bra{\chi_j(r)}\right)\\
 &=p\ket\eta\bra\eta+\frac{1-p}{3(1+r^2)}\sum_{j=0}^2
 \Bigl[r^2\ket{a_j}\bra{a_j}+\ket{b_j}\bra{b_j}\\
 &\hspace{43mm}+r\gamma\bigl(\ket{a_j}\bra{b_j}+\ket{b_j}\bra{a_j}\bigr)\Bigr].
\end{aligned}
\label{eq:family}
\end{equation}
\endgroup
Here $(a_0,b_0)=(00,12)$, $(a_1,b_1)=(11,20)$ and $(a_2,b_2)=(22,01)$, so that $\ket{\chi_j(r)}=(r\ket{a_j}+\ket{b_j})/\sqrt{1+r^2}$; only the coherences between $a_j$ and $b_j$ are damped, by the factor $\gamma$. We can also obtain the time-evolved density matrix simply by plugging our initial state \eqref{eq:mf} in above general solution. We denote the symmetric slice $r=1$ by $\rho_p(\gamma)$. The state family is defined here by \eqref{eq:kets}--\eqref{eq:family}. In our previous study on this problem, we had taken a mixture between state $\ket{\eta}$ and another state, which is entirely different from this construction \cite{Ali-MPLA-2019}. The construction is motivated by the stationary maximally entangled component used in that earlier study. Here its weight is explicit, and the negativity formula below determines precisely when a positive time-invariant value occurs. All three added amplitude ratios have the same cyclic orientation. Both marginals of \eqref{eq:family} are $I_3/3$: each local basis value receives diagonal weight $b+A+B=1/3$, with the coefficients defined below. In particular the states have full local support.

The state family belongs to the diagonal orthogonal invariant framework of Singh and Nechita~\cite{SinghNechita} and its cyclic sign invariant subclass. Its complete PPT/separability classification is an exact specialization of Gulati, Nechita and Singh~\cite[Theorem VI.5, arXiv v2]{GulatiNechitaSingh}. The static limiting simplex was already classified by Yu and Liu~\cite{YuLiu} and Qi and Hou~\cite{QiHou}; it includes the activation family of Horodecki, Horodecki and Horodecki~\cite{Horodecki} and is also studied by Ali~\cite{Ali2026}. The rank-four PPT-invariant initial curve is the family constructed by Ha, Kye and Park~\cite{HaKyePark}. These identifications fix the starting point of the present study.

We develop an explicit constructive and dynamical account of this class. A finite product decomposition, the two cyclic Choi witnesses~\cite{QiHou}, and a partial-transpose witness give complete certificates. A closed realignment formula~\cite{Rudolph,ChenWu} detects every PPT-entangled member and is constant along each PPT portion. We compute the negativity and transition times and show that one fixed two-level local projection certifies distillability throughout the NPT region. Finite-time loss of distillability under qutrit dephasing was established by Song, Chen and Zhu~\cite{Song2009} and Ali~\cite{Ali2010}; our results give an endpoint-complete analytic phase portrait for the specified trajectories. We also resolve their channel interpretation: the printed global dephasing map and a correlated local random-unitary semigroup agree on this family, although the former can entangle a product input and the latter cannot. In a dephasing trajectory, a vanishing negativity therefore does not decide whether entanglement has disappeared or has become PPT. An exact dynamical description requires both a separability certificate and an entanglement test beyond partial transposition.

\subsection{Complete PPT and separability regions}
We define 
\begin{equation}
 b=\frac p3,\qquad A=\frac{(1-p)r^2}{3(1+r^2)},\qquad
 B=\frac{1-p}{3(1+r^2)},\qquad c=\frac{(1-p)r\gamma}{3(1+r^2)}.
 \label{eq:coefficients}
\end{equation}
In the order $(02,10,21)$, $(00,12)$, $(11,20)$, $(22,01)$, the ordinary and partially transposed states are
\begin{equation}
\begin{aligned}
 \rho_{p,r}(\gamma)&\cong bJ_3\oplus
 \begin{pmatrix}A&c\\c&B\end{pmatrix}^{\oplus3},\\
 \rho_{p,r}(\gamma)^{T_B}&\cong[(b-c)I_3+cJ_3]\oplus
 \begin{pmatrix}A&b\\b&B\end{pmatrix}^{\oplus3}.
\end{aligned}
 \label{eq:blocks}
\end{equation}
Here $J_3$ is the all-ones matrix. The first line is positive because $c^2=AB\gamma^2\le AB$; its trace is $3b+3A+3B=1$.

The local permutation $V\ket j=\ket{j+1\bmod3}$ on $B$ gives the useful coefficient representation
\begin{equation}
 X_c=(I\otimes V)\rho_{p,r}(\gamma)(I\otimes V^\dagger)
 =bH_0+AD_+^{\prime}+BD_-^{\prime}+cC_{\rm sw},
 \label{eq:shifted}
\end{equation}
where, with indices modulo three,
\[
 H_0=\sum_{i,j}\ket{ii}\bra{jj},\quad D_0=\sum_i\ket{ii}\bra{ii},\quad
 D_\pm^{\prime}=\sum_i\ket{i,i\pm1}\bra{i,i\pm1},\quad
 C_{\rm sw}=\sum_{i<j}(\ket{ij}\bra{ji}+\ket{ji}\bra{ij}).
\]
This is a local diagonal orthogonal invariant state in Singh--Nechita's notation~\cite[Theorem 2.9 and Lemma 2.12]{SinghNechita}, with coefficient matrices
\[
 \mathsf A=\begin{pmatrix}b&A&B\\B&b&A\\A&B&b\end{pmatrix},\qquad
 \mathsf B=bJ_3,\qquad \mathsf C=(b-c)I_3+cJ_3.
\]
Its cyclic symmetry permits direct application of Gulati--Nechita--Singh~\cite[Theorem VI.5, arXiv v2]{GulatiNechitaSingh}: the circulant vectors are $(b,A,B)$, $(b,b,b)$ and $(b,c,c)$, up to reversal of the last two entries under the opposite cyclic convention. For $b>0$, $bJ_3$ generates an extreme ray of the circulant positive-semidefinite cone, and $\mathsf C\succeq0$ exactly when $c\le b$. Their theorem gives PPT iff $AB\ge b^2$, and separability iff $A,B\ge b$ in that sector. The remaining $c>b$ sector is NPT by \eqref{eq:blocks}; at $b=0$, separability occurs exactly when $c=0$. Thus the following is an explicit specialization of that classification. We give a self-contained constructive proof and detection formula.

\begin{theorem}
\label{thm:classification}
Let $0\le p,\gamma\le1$ and $r>0$, and define
\begin{equation}
 L_r(\gamma)=\frac{r\gamma}{1+r^2+r\gamma},\qquad
 U_r=\frac{r}{1+r+r^2},\qquad
 S_r=\frac{\min\{1,r^2\}}{1+r^2+\min\{1,r^2\}}.
 \label{eq:thresholds}
\end{equation}
Then
\begin{align}
 \rho_{p,r}(\gamma)\text{ is PPT}&\quad\Longleftrightarrow\quad L_r(\gamma)\le p\le U_r,
 \label{eq:ppt}\\
 \rho_{p,r}(\gamma)\text{ is separable}&\quad\Longleftrightarrow\quad L_r(\gamma)\le p\le S_r.
 \label{eq:separable}
\end{align}
An interval with its lower endpoint greater than its upper endpoint is empty. Among PPT states, the realignment criterion detects exactly those with $p>S_r$. For $r\ne1$, $S_r<U_r$, and this gives the whole PPT-entangled region.
\end{theorem}
\begin{proof}
The second line of \eqref{eq:blocks} is positive exactly when $b\ge c$ and $AB\ge b^2$. These inequalities are $p\ge L_r(\gamma)$ and $p\le U_r$. In particular $p\le U_r\le1/3$ for every PPT state.

We first give the separable decomposition. Put
\[
 H=3\ket\eta\bra\eta,\qquad
 D_+=\sum_{i=0}^2\ket{ii}\bra{ii},\qquad
 D_-=\ket{12}\bra{12}+\ket{20}\bra{20}+\ket{01}\bra{01}.
\]
Let $C$ be the sum of the six off-diagonal matrix units coupling the pairs $(00,12)$, $(11,20)$, $(22,01)$. Then $\rho_{p,r}(\gamma)=bH+AD_++BD_-+cC$. For $\zeta=\sqrt{-1}$ and $k,l\in\{0,1,2,3\}$, let
\[
 q_{kl}=(1,\zeta^k,\zeta^l)^T\otimes(\zeta^{-k},\zeta^{-l},1)^T,
 \quad Q_{16}=\sum_{k,l=0}^3q_{kl}q_{kl}^{\dagger},\quad
 Q_4=\sum_{k,l\in\{0,2\}}q_{kl}q_{kl}^{\dagger}.
\]
The phase characters give $Q_{16}/16=H+D_++D_-$ and $Q_4/4=H+D_++D_-+C$. Explicitly, the nine exponent pairs of $q_{kl}$ are
\[
 (-1,0),(0,-1),(0,0),(0,0),(1,-1),(1,0),(-1,1),(0,0),(0,1).
\]
Averaging modulo four keeps only the coincident constant characters in the support of $\eta$; averaging over even $k,l$ also keeps the three pair coherences. Consequently,
\begin{equation}
 \rho_{p,r}(\gamma)=(b-c)\frac{Q_{16}}{16}+c\frac{Q_4}{4}
 +(A-b)D_++(B-b)D_-.
 \label{eq:product}
\end{equation}
Every summand is a sum of product projectors. The coefficients are nonnegative when $b\ge c$ and $b\le\min\{A,B\}$, exactly the region \eqref{eq:separable}. Each $q_{kl}$ has squared norm nine, so the normalized weights in \eqref{eq:product} sum to $9(b-c)+9c+3(A-b)+3(B-b)=1$. These averages are finite versions of the diagonal unitary and orthogonal averages in Singh and Nechita~\cite[Proposition 2.3]{SinghNechita}.

To prove the converse within the PPT region, define realignment by $\R(\rho)_{ik,jl}=\rho_{ij,kl}$. A separable density matrix satisfies $\|\R(\rho)\|_1\le1$~\cite[Section 2]{ChenWu}. For completeness, a normalized product projector realigns to a rank-one matrix of trace norm one, and convexity gives this inequality for a separable mixture.

Apply the local permutation $\ket{j}_B\mapsto\ket{j+1\bmod3}_B$, taking $\eta$ to $(\ket{00}+\ket{11}+\ket{22})/\sqrt3$. Realignment, up to row and column permutations, is the direct sum of
\begin{equation}
 P=\begin{pmatrix}b&A&B\\B&b&A\\A&B&b\end{pmatrix},\qquad
 \begin{pmatrix}b&c\\c&b\end{pmatrix}^{\oplus3}.
 \label{eq:realignment-blocks}
\end{equation}
Local unitaries preserve the realignment singular values. The normal circulant matrix $P$ has singular values $1/3$ and
$\sqrt{(b-(A+B)/2)^2+3(A-B)^2/4}$, the latter twice. The remaining blocks have trace norm $2\max\{b,c\}$ each. Thus, on the whole parameter domain,
\begin{equation}
 \|\R(\rho_{p,r}(\gamma))\|_1=
 \frac13+\frac13\sqrt{(3p-1)^2+3(1-p)^2\delta_r^2}
 +6\max\{b,c\},\qquad \delta_r=\frac{r^2-1}{r^2+1}.
 \label{eq:realignment}
\end{equation}
In the PPT region $b\ge c$, so the last term is $2p$. Since $p\le1/3$, the strict inequality $\|\R(\rho)\|_1>1$ is equivalent to
\[
 \sqrt{(1-3p)^2+3(1-p)^2\delta_r^2}>2(1-3p)
 \quad\Longleftrightarrow\quad
 (1-p)|\delta_r|>1-3p.
\]
For $r\ge1$ this is $p>1/(r^2+2)$, and for $0<r\le1$ it is $p>r^2/(1+2r^2)$. These are precisely $p>S_r$. Every PPT state outside the product region is therefore entangled. Outside the PPT region the state is entangled by partial transposition, completing the classification.
\end{proof}

\subsection{Fixed witnesses and local dephasing realization}
The classification also has a complete description using three fixed witnesses. In the shifted basis let
\begin{equation}
 W_\pm=\frac{2D_0+D_\pm^{\prime}-H_0}{3},\qquad
 W_0=(\ket v\bra v)^{T_B},\quad v=(\ket{00}-\ket{11})/\sqrt2.
 \label{eq:fixed-witnesses}
\end{equation}
The first two are the witnesses associated with the two cyclic Choi maps~\cite[Theorem 3.2]{QiHou},
\[
 (\Psi_\pm(Z))_{ii}=z_{ii}+z_{i\pm1,i\pm1},\qquad
 (\Psi_\pm(Z))_{ij}=-z_{ij}\quad(i\ne j).
\]
Specifically $W_\pm=(\id\otimes\Psi_\mp)(H_0/3)$. To check positivity of the maps on $xx^\dagger$, write $a_i=|x_i|^2$. For positive $a_i$, the output is $\operatorname{diag}(2a_i+a_{i\pm1})-xx^\dagger$, which is positive iff $\sum_i a_i/(2a_i+a_{i\pm1})\le1$. For the plus orientation, clearing positive denominators reduces this inequality to
$a_0^2a_2+a_0a_1^2+a_1a_2^2\ge3a_0a_1a_2$, by AM--GM. The minus orientation reverses the indices, and zero coordinates follow by continuity. Positivity on rank-one inputs implies positivity on every positive matrix. The witness $W_0$ is block positive because $\langle x,y|W_0|x,y\rangle=|\langle v|x,\overline y\rangle|^2$.

Direct substitution in \eqref{eq:shifted} gives
\begin{equation}
 \Tr(W_0X_c)=b-c,\qquad \Tr(W_+X_c)=A-b,\qquad \Tr(W_-X_c)=B-b.
 \label{eq:witness-values}
\end{equation}
Consequently $X_c$ is separable exactly when these three expectations are nonnegative. In particular a fixed negative $W_+$ or $W_-$ expectation certifies the persistent PPT-entangled tail. The maps and witnesses are established ones; \eqref{eq:witness-values} is their explicit application to this family.

A local dephasing model realizes exactly the same trajectories. Put $D_{kl}=\operatorname{diag}(1,i^k,i^l)$, $k,l=0,1,2,3$, and define
\begin{equation}
 \mathcal T(Z)=\frac1{16}\sum_{k,l}(D_{kl}\otimes\overline D_{kl})Z
 (D_{kl}\otimes\overline D_{kl})^\dagger,\qquad
 \mathcal F_\gamma=\gamma\,\id+(1-\gamma)\,\mathcal T.
 \label{eq:local-channel}
\end{equation}
The phase average fixes $H_0,D_+^{\prime},D_-^{\prime}$ and removes $C_{\rm sw}$. Since $\mathcal T^2=\mathcal T$,
\[
 \mathcal F_{\gamma_1}\mathcal F_{\gamma_2}=\mathcal F_{\gamma_1\gamma_2},\qquad
 \mathcal F_\gamma(X_{\sqrt{AB}})=X_{\gamma\sqrt{AB}}.
\]
Thus $\gamma=e^{-\Gamma t/2}$ gives a semigroup generated by $(\Gamma/2)(\mathcal T-\id)$. Each constituent unitary in \eqref{eq:local-channel} is a product unitary, chosen using shared classical randomness. For $U=I\otimes V$, the channel $\operatorname{Ad}(U^\dagger)\circ\mathcal F_\gamma \circ\operatorname{Ad}(U)$ returns the trajectories in the original basis, where $\operatorname{Ad}(U)(Z)=UZU^\dagger$.

The full printed channel \eqref{Eq:MF} has a different action on general inputs. Its generator is $-(\Gamma/2)[Q,[Q,\cdot]]$, $Q=\operatorname{diag}(h)$. The nonadditive charge array precludes identifying this generator with the additive local Hamiltonian preceding Eq.~(4) in Ali~\cite{Ali-MPLA-2019}. More explicitly, put $s=(\ket0+\ket1+\ket2)/\sqrt3$ and $\sigma_\gamma=\mathcal E_\gamma(\ket{s,s}\bra{s,s})$. The principal submatrix of $\sigma_\gamma^{T_B}$ on $(00,01,21)$ is
\begin{equation}
 \frac19\begin{pmatrix}1&\gamma&1\\\gamma&1&1\\1&1&1\end{pmatrix},\qquad
 \det=-\frac{(1-\gamma)^2}{729}<0\quad(0\le \gamma<1).
 \label{eq:channel-obstruction}
\end{equation}
It therefore entangles this product input. The global channel is valid and agrees with \eqref{eq:local-channel} on the specified state trajectories after the basis change, but it cannot be a mixture of local product-unitary operations on arbitrary inputs.

\subsection{Negativity and exact dynamical regimes}
The complete partial-transpose spectrum read from \eqref{eq:blocks} is
\begin{equation}
\begin{gathered}
 b+2c,\qquad (b-c)^{\times2},\\
 \left(\frac{A+B+\sqrt{(A-B)^2+4b^2}}2\right)^{\!\times3},\qquad
 \left(\frac{A+B-\sqrt{(A-B)^2+4b^2}}2\right)^{\!\times3}.
\end{gathered}
 \label{eq:pt-spectrum}
\end{equation}
With $[x]_+=\max\{x,0\}$ and the negativity convention
$N(\rho)=(\|\rho^{T_B}\|_1-1)/2$~\cite{VidalWerner}, this gives
\begin{equation}
 N(\rho_{p,r}(\gamma))=
 \frac32\left[\sqrt{(A-B)^2+4b^2}-(A+B)\right]_+
 +2[c-b]_+.
 \label{eq:general-negativity}
\end{equation}
Only the second term depends on time. If $p>U_r$, then $b>\sqrt{AB}\ge c$, so the negativity is positive and constant for the whole trajectory. For $p<1$ the state is nonstationary, since $c$ decays strictly with time. At $p=1$ it is the stationary pure state $\ket\eta\bra\eta$. If $0<p\le U_r$, the first term vanishes, and the second is nonincreasing and vanishes from the PPT onset onward. If $p=0$, negativity equals $2r\gamma/[3(1+r^2)]$ and vanishes only asymptotically. Thus positive time-invariant negativity occurs exactly when $p>U_r$.

\begin{proposition}
Every NPT state in \eqref{eq:family} is one-copy distillable.
\end{proposition}
\begin{proof}
In the shifted basis project each subsystem onto levels $\{0,1\}$. The unnormalized two-qubit output and its partial transpose are
\[
 Y=\begin{pmatrix}b&0&0&b\\0&A&c&0\\0&c&B&0\\b&0&0&b\end{pmatrix},\qquad
 Y^{T_B}=\begin{pmatrix}b&0&0&c\\0&A&b&0\\0&b&B&0\\c&0&0&b\end{pmatrix}.
\]
The success probability is $2b+A+B=(1+p)/3>0$. The partial transpose has a negative eigenvalue exactly when $c>b$ or $b^2>AB$, the entire NPT region of \eqref{eq:blocks}. The projection therefore yields an entangled two-qubit state in every NPT case, establishing one-copy distillability~\cite{HorodeckiBound}. This means distillability following a projection on one copy, rather than extraction of a perfect singlet from one copy.
\end{proof}

For fixed $r>0$ and $0<p\le U_r$, the state becomes PPT when $c=b$, at
\begin{equation}
 \gamma_{\mathrm P}=\frac{p(1+r^2)}{r(1-p)},\qquad
 t_{\mathrm P}=\frac2\Gamma\log\frac{r(1-p)}{p(1+r^2)}.
 \label{eq:ppt-time}
\end{equation}
If $S_r<p<U_r$, then $0<\gamma_{\mathrm P}<1$: the state is initially NPT and one-copy distillable, becomes PPT entangled at $t_{\mathrm P}$, and remains PPT entangled for every later time, including the asymptotic limit. This is a distillability-sudden-death trajectory with a persistent bound-entangled tail. This dynamics is also remarkable as in all previous studies of distillability sudden, it is was observed that states ultimately become separable. In our case, we have situation where the states end as bound entangled states. At $p=U_r$ the state is PPT from $t=0$ and remains PPT entangled for $r\ne1$. If $0<p\le S_r$, it is separable once it is PPT; the finite transition is entanglement sudden death. When $r=1$ and $p=S_r=U_r=1/3$, the state is separable from $t=0$. Throughout any PPT portion, \eqref{eq:realignment} is independent of $\gamma$. This is a constant detection norm; it does not assert constancy of an entanglement measure.

\subsection{The symmetric slice and its initial boundary}
For $r=1$, put $a=(1-p)/6$, $b=p/3$. Partial transposition gives
\begin{equation}
 \rho_p(\gamma )^{T_B}\cong
 \begin{pmatrix}a&b\\b&a\end{pmatrix}^{\oplus3}
 \oplus\bigl[(b-a\gamma)I_3+a\gamma J_3\bigr],
 \label{eq:native-pt}
\end{equation}
where $J_3$ is the all-ones matrix, the pairs are $(00,12)$, $(11,20)$, $(22,01)$, and the last block is on $(02,10,21)$. Thus, with 
$[x]_+=\max\{x,0\}$,
\begin{equation}
 N(\rho_p(\gamma ))=\frac{[3p-1]_+}{2}+
\frac{[(1-p)\gamma -2p]_+}{3},\qquad
 N(\rho)=\frac{\|\rho^{T_B}\|_1-1}{2}.
\label{eq:negativity}
\end{equation}
In particular, the positive map $T(X)=X^{\mathsf T}$ detects the symmetric states with
\begin{equation}
 p>\frac13\quad\text{or}\quad p<\frac{\gamma }{\gamma +2}.
 \label{eq:transpose-range}
\end{equation}
For the second range, the fixed vector $v=\ket{02}-\ket{10}$ gives the explicit certificate
\begin{equation}
 W_v=(\ket v\bra v)^{T_B},\qquad
 \Tr(W_v\rho_p(\gamma ))=\frac{2p-(1-p)\gamma }{3}<0.
 \label{eq:native-witness}
\end{equation}
Indeed $\langle x,y|W_v|x,y\rangle=|\langle v|x,\overline y\rangle|^2\ge0$. For $p>1/3$, a difference vector in any of the two-dimensional blocks in \eqref{eq:native-pt} gives a negative expectation. This is the partial-transpose criterion~\cite{Peres}.

For $0<p<1/3$, entanglement disappears at
\begin{equation}
 t_*(p)=\frac2\Gamma\log\frac{1-p}{2p}.
 \label{eq:native-death}
\end{equation}
Theorem~\ref{thm:classification} proves separability from that time onward. This is entanglement sudden death. At $p=0$, entanglement persists at every finite time and vanishes only in the limit. For $1/3<p<1$, negativity is positive and time-invariant along a nonstationary trajectory; $p=1$ is stationary. Transposition therefore detects every entangled state of the symmetric family, including the initial states below $p=1/3$; after the stated transition those states are separable.

At the initial boundary the decomposition can be shortened to four normalized product states. For $\epsilon,\delta\in\{1,-1\}$, put
\[
 u_{\epsilon\delta}=(1,\epsilon,\delta)^T/\sqrt3,\qquad
 v_{\epsilon\delta}=(\epsilon,\delta,1)^T/\sqrt3.
\]
Then
\begin{equation}
 \rho_{1/3,1}(1)=\frac14\sum_{\epsilon,\delta=\pm1}
 |u_{\epsilon\delta}\rangle\langle u_{\epsilon\delta}|\otimes
 |v_{\epsilon\delta}\rangle\langle v_{\epsilon\delta}|.
 \label{eq:four-product}
\end{equation}
Indeed this is $Q_4/36$, which equals the state by \eqref{eq:product} at $b=A=B=c=1/9$. The four mutually orthogonal initial supports have nonzero eigenvalues $1/3,2/9,2/9,2/9$, so the state has rank four. Every decomposition into pure product projectors therefore needs at least four terms, proving minimality.

\section{Exact examples and a boundary curve}
\subsection{Known static families and their embeddings}
In the shifted basis the limit $\gamma =0$ is
\begin{equation}
 p\ket\Omega\bra\Omega+q_+\sigma_++q_-\sigma_-,\qquad
 q_+=\frac{(1-p)r^2}{1+r^2}=3A,\quad q_-=\frac{1-p}{1+r^2}=3B,
 \label{eq:ali-slice}
\end{equation}
where $\Omega=(\ket{00}+\ket{11}+\ket{22})/\sqrt3$ and $\sigma_\pm=D_\pm^{\prime}/3$. Every mixture with positive noise weights is represented; the zero-weight edges follow as limits $r\to0$ or $r\to\infty$. This is the static simplex classified by Yu--Liu~\cite[Eq.~(16) and the discussion after Eq.~(17)]{YuLiu} and Qi--Hou~\cite[Example 3.3]{QiHou}, and considered by Ali~\cite{Ali2026}. With $a=p$, its known classification is
\[
 \mathrm{PPT}\iff q_+q_-\ge a^2,\qquad
 \mathrm{SEP}\iff a\le\min(q_+,q_-).
\]
The two cyclic map orientations detect the two PPT-entangled regions $a>q_-$ and $a>q_+$, respectively. Equation~\eqref{eq:product} supplies a finite preparation certificate for the complementary triangle. For $p=2/7$, the activation family of Horodecki--Horodecki--Horodecki~\cite{Horodecki} is recovered with $\alpha=5r^2/(1+r^2)$; in particular $r=2$ gives $\alpha=4$.

\subsection{A rational transition benchmark}
For an explicit rational example take $r=2$, $p=1/4$. The blocks in \eqref{eq:blocks} have
\begin{equation}
 b=\frac1{12},\qquad A=\frac15,\qquad B=\frac1{20},\qquad c=\frac \gamma {10}.
 \label{eq:example}
\end{equation}
At $\gamma =1$, the partial transpose has two negative eigenvalues $-1/60$, and its other eigenvalues are positive, giving $N=1/30$. The exact transition is
\begin{equation}
 \gamma _{\mathrm P}=\frac56,\qquad t_{\mathrm P}=\frac2\Gamma\log\frac65.
 \label{eq:example-time}
\end{equation}
For every $0\le \gamma \le5/6$,
\begin{equation}
 \|\R(\rho_{1/4,2}(\gamma ))\|_1=\frac{25+\sqrt{67}}{30}>1.
 \label{eq:example-ccnr}
\end{equation}
The inequality follows from $67>25$. At $\gamma =1/2$, corresponding to $t=2\log2/\Gamma$, its partial transpose is positive definite. The exact spectra are
\begin{equation}
\begin{aligned}
 \spec\rho_{1/4,2}(1/2)&=\left\{\frac14,0,0,
 \left(\frac{5+\sqrt{13}}{40}\right)^{\!\times3},
 \left(\frac{5-\sqrt{13}}{40}\right)^{\!\times3}\right\},\\
 \spec\rho_{1/4,2}(1/2)^{T_B}&=\left\{\frac{11}{60},
 \left(\frac1{30}\right)^{\!\times2},
 \left(\frac{15+\sqrt{181}}{120}\right)^{\!\times3},
 \left(\frac{15-\sqrt{181}}{120}\right)^{\!\times3}\right\}.
\end{aligned}
 \label{eq:example-spectra}
\end{equation}
The realignment singular values are $1/3$, $\sqrt{67}/60$ twice, $2/15$ three times, and $1/30$ three times. Thus \eqref{eq:example} supplies a rational matrix with explicit certificates of positivity, positive definiteness of its partial transpose, and entanglement.

\subsection{Dephasing the Ha--Kye--Park curve}
The rank-four initial curve has an exact established identification. With $D_r=r^2+r+1$, $p_r=r/D_r=U_r$, $\Phi=\sum_i\ket{ii}$ and $z_i=\sqrt r\ket{i,i+1}+r^{-1/2}\ket{i+1,i}$,
\[
 X_{\sqrt{AB}}=\frac{\ket\Phi\bra\Phi+\sum_i\ket{z_i}\bra{z_i}}
 {3(1+r+r^{-1})}\quad(p=p_r).
\]
This is the PPT-invariant family of Ha--Kye--Park~\cite[Section 4]{HaKyePark}, with their parameters $\lambda=\sqrt r$, $\mu=r^{-1/2}$; for $r\ne1$ they prove PPT entanglement and extremality. For the associated trajectories, $\gamma _{\mathrm P}=1$, and
\begin{equation}
 \rho_{p_r,r}(1)^{T_B}=\rho_{p_r,r}(1),\qquad
 \rho_{p_r,r}(1)\longrightarrow\rho_{1/3,1}(1)\quad(r\to1).
 \label{eq:close}
\end{equation}
These states are PPT entangled for every $r\ne1$. Indeed, their realignment gap, constant throughout the evolution, is
\begin{equation}
 \|\R(\rho_{p_r,r}(\gamma ))\|_1-1
 =\frac{2|r-1|\bigl(\sqrt{D_r}-|r-1|\bigr)}{3D_r}>0,
 \label{eq:close-gap}
\end{equation}
since $D_r-(r-1)^2=3r>0$. Their rank at the initial time is four. For $r\ne1$ the trajectory is nonstationary, and remains PPT entangled at every time. The reciprocal asymmetric static endpoint at $\gamma =0$ and its realignment detection are also treated by Johnston and MacLean~\cite[Examples 1 and 5]{JohnstonMacLean}.

\section{Conclusions}
The common-ratio family separates the mechanisms behind time-invariant negativity and the loss of distillability. For $U_r<p<1$, the state changes with time while its negativity remains strictly positive and constant. For $0<p\le U_r$, the decaying coherence reaches the PPT boundary at the explicit time \eqref{eq:ppt-time}: the subsequent state is separable when $p\le S_r$, and PPT entangled when $p>S_r$. At $p=0$, negativity vanishes only asymptotically; at $p=1$, the state itself is stationary. The symmetric slice $r=1$ has no PPT-entangled region, whereas an amplitude imbalance allows a persistent bound-entangled tail.

The product decomposition, realignment formula and fixed witnesses make the established cyclic-state classification constructive. A single fixed local projection covers the entire NPT region, so the phase portrait distinguishes genuine disentanglement from the conversion of distillable entanglement to bound entanglement. The rational example and the exact Horodecki and Ha--Kye--Park embeddings provide explicit checks and connect the trajectories to established state families.

The channel distinction is essential to the physical interpretation. The printed entrywise map can entangle a product input, while the correlated local random-unitary semigroup cannot; their agreement holds on the specified trajectories rather than on arbitrary states. The latter semigroup therefore supplies a local dephasing realization of the phase portrait. Extending the classification to three independently chosen amplitude ratios, or to a nonextremal maximally correlated block, remains a separate problem.

\subsection*{Exact verification}
All-parameter claims follow from the displayed proofs. Two separately implemented symbolic programs reconstruct the state from its initial kets and verify the channel factors, both marginals, characteristic polynomials, finite product identity and realignment Gram polynomial using exact arithmetic. A further exact program checks the fixed witnesses, the local twirl and semigroup, the product-input obstruction, and the two-qubit projection. The programs also check the rational spectra in \eqref{eq:example-spectra}, the transition value $\gamma =5/6$, and the Horodecki identification. The calculations corroborate the proofs and can be reproduced from the accompanying source code; no numerical sampling or Lean formalization is used to establish this classification.

\subsection*{Source and code availability}
The accompanying source package contains the editable manuscript and the four exact-arithmetic verification programs, with pinned Python dependencies and reproduction instructions.

\subsection*{Funding}
This research received no specific funding.

\subsection*{Competing interests}
The authors declare no competing interests.

\subsection*{Use of AI tools}
Haobo Ma and Wenlin Zhang used AI tools, namely ChatGPT Pro (GPT-6 Pro), Codex (gpt-6.1-sol) and Claude Opus 5.5 (via Claude Code), for literature search, computation and double-checking. All proofs can be checked by hand, and the computations can be reproduced from the verification package. H. Ma and W. Zhang take full responsibility for their contributions.

\end{document}